\documentclass{kluwer}    
\usepackage{graphicx}

\newdisplay{guess}{Conjecture}

\begin{document}                                                                                   
\begin{article}
\begin{opening}         
\title{The orbital period distribution of subdwarf B binaries} 
\author{Luisa \surname{Morales-Rueda} \email{lmr@astro.soton.ac.uk}}  
\runningauthor{L. Morales-Rueda}
\runningtitle{Orbital period distribution of sdB binaries}
\institute{University of Southampton}

\author{Pierre F. L. \surname{Maxted} \email{pflm@astro.keele.ac.uk}}  
\institute{Keele University}

\author{Tom R. \surname{Marsh} \email{T.R.Marsh@warwick.ac.uk}}  
\institute{Warwick University}


\begin{abstract}
We present the results of a 3.5 year long campaign to measure orbital
periods of subdwarf B (sdB) star binaries. We directly compare our
observed orbital period distribution with that predicted by using
binary population synthesis. Up to now, most of our systems seem to
have been formed through two of the formation channels discussed by
Han et al.(2003), i.e. the first and the second common envelope
ejection (CE) channels. At this point, thanks to the long baseline of
our observations, we are starting to detect also very long orbital
period systems. These have probably come from a complete different
formation path, the first stable Roche Lobe overflow (RLOF) channel in
which the first mass transfer phase is stable. This channel is
expected to lead to the formation of very wide binaries with typical
orbital periods ranging from one month to one year.
\end{abstract}

\end{opening}           

\section{Introduction}  
The gravities and temperatures of subdwarf B stars (sdB) suggest that
they are composed of a helium core of 0.5\,M$_\odot$ and a very thin
hydrogen layer of $\lsim$0.02\,M$_\odot$. Although this has been known
for some years it is still unclear how these stars lose most of their
hydrogen envelopes and still manage to ignite helium. Both, single
star and binary star evolution have been suggested as possible
formation channels. In 2001, Maxted et al. found that a high fraction
of sdB stars from the PG survey, 2/3, are short period binaries. The
other third is thought to be made of long period binaries that formed
through stable Roche Lobe overflow (i.e. no common envelope) and
single stars (Green, Liebert \&\ Saffer 2001). These results indicate
that formation via a binary evolution is one of the main channels to
make sdB stars.

In this paper we present new results of the systematic search for sdB
binary systems that we started a few years ago (Maxted et al. 2001,
Morales-Rueda et al. 2003). We also make use of the fact that short
period sdB binaries do not change their orbital periods significantly
after they emerge from the common envelope phase to test models of
binary evolution. We compare our results with those obtained by using
population synthesis models by Han et al. (2002, 2003).

\section{Description of the sample and techniques}
We started this program in April 2000 and have been collecting data
since then. Most of the spectra have been obtained with the
intermediate resolution spectrograph IDS on the 2.5\,m INT telescope
and only in the last year we have started collecting spectra for
Southern Hemisphere targets with the 1.9\,m Radcliffe telescope at
SAAO. The spectra taken on the first year of the program covered the
H$\alpha$ line. From August 2001 we obtained blue spectra instead,
covering the Balmer lines from H$\beta$ to H$\epsilon$ in most
cases. By taking blue spectra we can not only look for radial velocity
variations in the lines and measure orbital periods of the sdB
binaries, but we can also fit the profiles of the lines by a grid of
synthetic spectra and measure their effective temperatures, surface
gravities and helium abundances (Saffer et al. 1994; Napiwotzki 1997;
Heber, Reid \&\ Werner 2000). A complete description of the data taken
with the INT is given in Morales-Rueda et al. (2003). Our sample (as
of July 2003) contains 117 systems, 80 from the Palomar-Green (PG)
survey, 16 from the Kitt Peak Downes (KPD) survey and 12 from the
Edinburgh-Cape (EC) survey.

\subsection{Radial velocity measurements}
\begin{figure*}
\begin{picture}(100,0)(10,20)
\put(0,0){\includegraphics{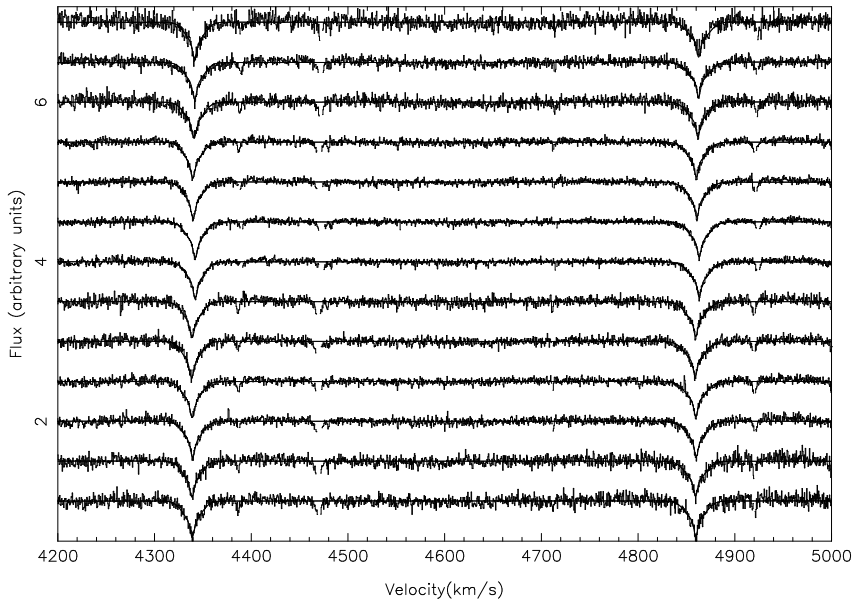}}
\noindent
\end{picture}
\vspace{80mm}
\caption{Example of least squares fitting to a set of 13 blue spectra
  of the sdB binary EC12327-1338. The fits shown are the ones obtained
  in the last fitting step described in the text. Both absorption
  lines are fitted simultaneously.}
  \label{fig1}
\end{figure*}

To measure the radial velocities of the lines we perform least squares
fitting of a model line profile that consists of 3 Gaussians. This
fitting consists of several steps: {\bf 1.} continuum normalise the
spectra, {\bf 2.} use default fixed values for the FWHMs and heights
of the 3 Gaussians and fit all the spectra simultaneously to obtain
the overall mean velocity, {\bf 3.} fit all the lines simultaneously
by fixing the overall velocity (obtained in previous step) and
allowing the FWHMs and heights of the 3 Gaussians to vary {\bf 4.}
fit the spectra individually this time fixing the FWHMs and heights to
the values obtained in step 3 and allowing for the velocities to be
variable, {\bf 5.} shift out the velocities obtained in step 4 for
each spectrum and obtain a more refined value for the FWHMs and
heights of the 3 Gaussians and {\bf 6.} fix the FWHMs and heights to
the refined values from step 5 and measure individual velocities for
each spectrum. We use as many Balmer lines as there are present in
each spectrum to perform the fitting. See Fig.~\ref{fig1} for an
example of fitting two Balmer lines.

\subsubsection{Uncertainties}
For the red spectra (up to August 2001) the statistical uncertainties
are of the order of 1\,km/s whereas for the blue spectra the
uncertainties are of the order of 2\,km/s (a tenth of a pixel). We
know that there are also other unaccounted sources of error that are
probably not correlated with the orbit or the statistical errors, i.e.
slit-filling, intrinsic variability of the star. We estimate these
systematic errors by assuming that when added in quadrature with the
raw uncertainties they will give us a reduced $\chi^2$ = 1. We assume
that in all cases the systematic uncertainty is at least 2\,km/s
($\sim$ a tenth of a pixel) and this value works well for most
systems. In some cases the systematic error turns out to be larger,
with the largest value being 5\,km/s for KPD0025+5402.

\subsection{Orbital period determination}

\begin{figure*}
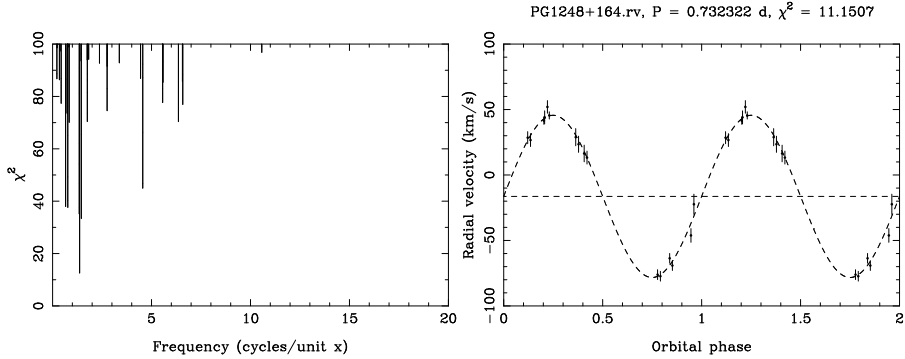

\begin{picture}(100,0)(10,20)
\put(0,0){\includegraphics{08052001pgramBW.ps}}
\put(0,0){\includegraphics{PG1248+164foldBW.ps}}
\noindent
\end{picture}
\vspace{50mm}
\caption{Left panel: example of a periodogram, $\chi^2$ versus
  frequency, for the sdB binary PG1248+164. Right panel: the data
  folded in the orbital period for the same sdB. These results are
  discussed in detail in Morales-Rueda et al. 2003.}
  \label{fig2}
\end{figure*}

We fit the data with a model composed of a sinusoid and a constant
(Cumming, Marcy \&\ Butler 1999) to determine the orbital
periods. This method works better than the Lomb-Scargle periodogram
for small numbers of points. For each fit we calculate its associated
$\chi^2$ and select the solution with the minimum $\chi^2$ value. An
example of a periodogram and a radial velocity curve folded on the
orbital period are presented in Fig.\ref{fig2}. 

To distinguish between competing aliases (those separated by
$\Delta\chi^2 \le$ 20) we use a technique developed by Marsh, Dhillon
\&\ Duck (1995) and later adapted to this study by Morales-Rueda et al.
(2003) to calculate the probability of the true orbital period being
further than 1 and 10 per cent from the value obtained. We assume our
values are correct if the probability of the period being wrong is
$\le$ 0.1 per cent.

\section{Results: The orbital period distribution}

\begin{figure*}[!ht]
\begin{picture}(100,0)(10,20)
\put(0,0){\includegraphics{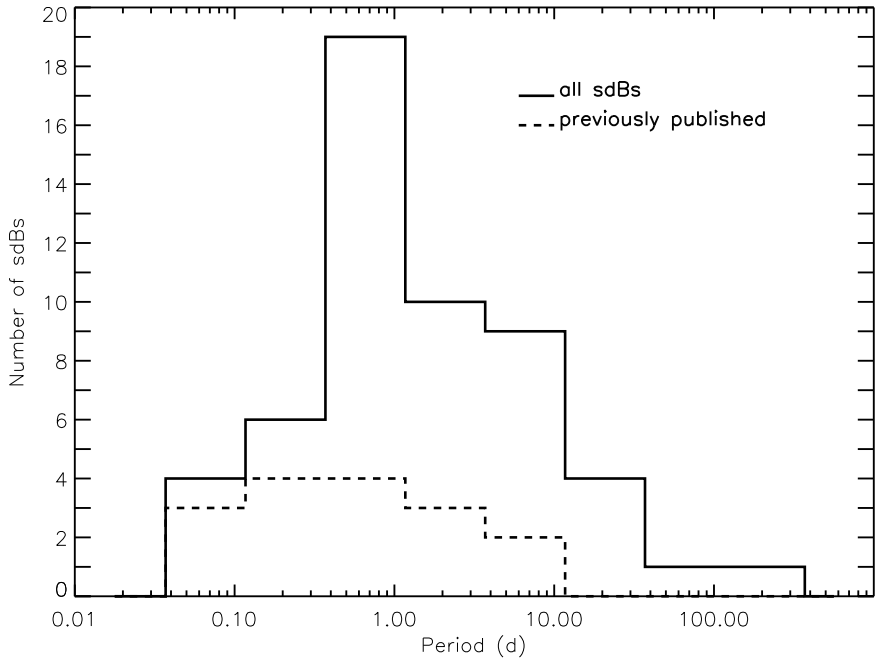}}
\noindent
\end{picture}
\vspace{90mm}
\caption{Orbital period distribution for the sdB binaries known.}
  \label{fig3}
\end{figure*}

In Fig.~\ref{fig3} we present the number of sdB binaries known when
this paper was produced versus their orbital periods. The systems
included in the figure are those presented in Fig. 4 of Morales-Rueda
et al. (2003) plus 7 new ones from our sample. During this workshop a
few more sdB binaries were presented by Edelmann et al. which have not
been included in the figure.

The picture we now have of how the orbital periods of sdB binaries are
distributed is very different from what it was before we started this
program. We can see that the distribution extends toward longer
periods than it was initially thought. Another important feature is
the excess of systems at orbital periods of the order of 1 day (the 1
day bin actually includes systems with orbital periods between 0.32
and 1 days). These systems should be more difficult to detect as one
requires more observations to be able to distinguish between the
orbital period and the observational aliases. This feature has been
seen by other authors (Green, private communication).

Thanks to the long time baseline of our observations we are starting
to detect binary systems with longer orbital periods, i.e. 4 systems
with orbital periods of tens of days and the first one with a period
of hundreds of days.

\subsection{Companions to the sdBs}

Maxted et al. discuss somewhere else in these proceedings the nature
of the companions to these sdBs in detail. It is worth pointing out
here that we find 2 systems in our sample with M dwarf companions,
PG1017-086 (Maxted et al. 2002) and PG1329+159 (Maxted et al. in
preparation). 19 others are found to have white dwarf (WD) companions
(Maxted et al. in preparation).

\section{Discussion: comparison with population synthesis models}

\begin{figure*}[!ht]
\begin{picture}(100,0)(10,20)
\put(0,0){\includegraphics{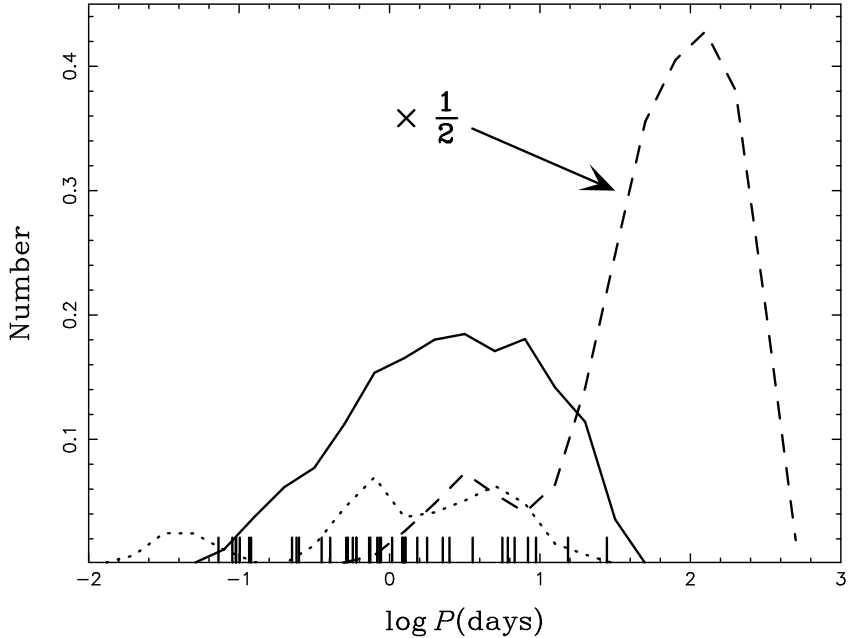}}
\noindent
\end{picture}
\vspace{85mm}
\caption{Model predictions from Han et al. 2003. The solid line
  represents the distribution of sdB binaries formed via the first CE
  ejection channel; the dotted line, those formed via the second CE
  ejection channel; and the dashed line the binaries formed through
  the first stable RLOF channel. The ticks at the bottom of the plot
  represent our data from Morales-Rueda et al. 2003.}
  \label{fig4}
\end{figure*}

Han et al. (2002, 2003) present five possible channels for the
formation of sdB stars, three of them will form sdBs in binary
systems. Fig.~\ref{fig4} shows the orbital period distribution
predicted by their favoured model. The plot also presents our data in
the form of vertical ticks on the horizontal axis. Most of the
binaries in our sample seem to have formed via the second common
envelope (CE) ejection channel (short period binaries with WD
companions), some via the first CE ejection channel (short to
intermediate periods with main sequence companions). Their models
predict the formation of a large number of long period binaries via
the first Roche Lobe overflow (RLOF) channel (first suggested by
Tutukov \&\ Yungleson 1990). Up to this day we have only detected one
of such system with an orbital period of a few hundred days in our
sample. We seem to be missing a very large population of binaries with
large orbital periods.

\section{How biased is our sample?}

\begin{figure*}[!ht]
\begin{picture}(100,0)(10,20)
\put(0,0){\includegraphics{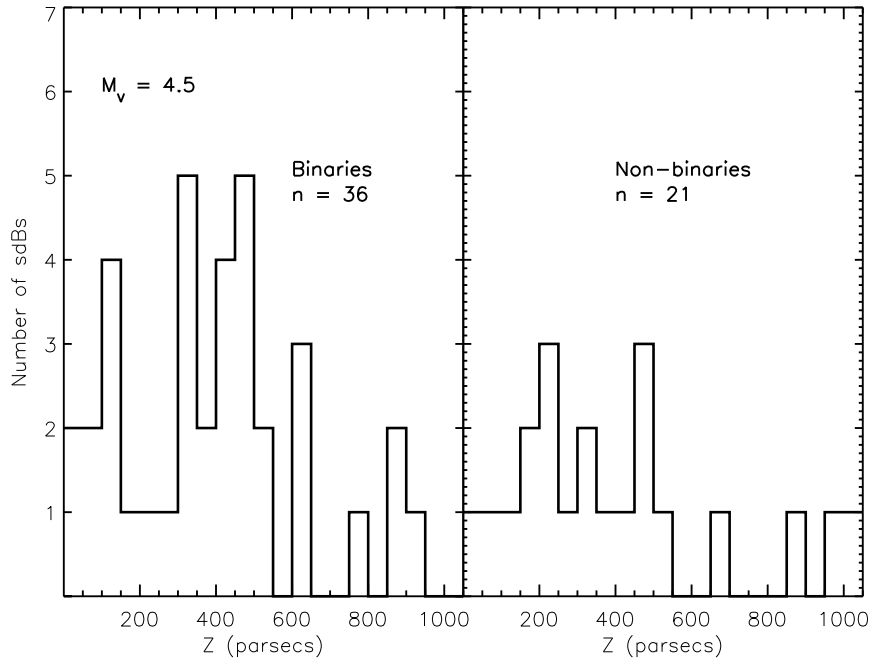}}
\noindent
\end{picture}
\vspace{90mm}
\caption{Distribution of the sdBs in our sample in the Galaxy. We
assume an absolute magnitude of 4.5 for an sdB star in order to
calculate their distances. All the systems in our sample are confined
to the disk of the Galaxy. We plot separately the systems in our
sample that we know are binaries from those that we think are not.
There is no significant difference in their distribution with scale
height. The 60 systems of our sample not included in this plot are
those for which we do not have enough observations yet to say whether
they are binaries or not.}
  \label{fig5}
\end{figure*}

Our sample suffers from several important biases:
\begin{itemize}
\item 1. Targets have been selected from 3 main surveys, PG, KPD and
EC. The PG survey is biased against companions of G and K spectral
type as any target with a spectrum showing CaII H-lines was taken off
the survey. This should be taken into account when looking at the
companions of sdBs as most of the ones from the PG survey will be WDs
instead of main sequence stars. This has important consequences when
one intends to compare binary formation models with observations. It
is difficult to assess how important this bias is.
\item 2. A second important bias is given by the resolution of the
spectra. Up to now we have used intermediate resolution spectrographs
to measure radial velocities. Longer period systems will show smaller
amplitude radial velocities and we are probably missing them. The use
of higher resolution spectrographs might change significantly the
picture we have at the moment.
\item 3. Only now, after almost four years of observations, is the time
baseline getting long enough to detect the longer period systems. This
can explain the dearth of long period systems compared to Han et al.'s
(2003) predictions.
\item 4. All targets we have observed are bright (V $<$ 14.5). It is
  also well known that the PG survey is incomplete at bright
  magnitudes due to saturation of the plates.
\item 5. All targets are in the disk of the Galaxy (see
Fig.~\ref{fig5}). 
\end{itemize}

\section{Conclusions}

The more sdBs we study the more binaries we find, the fraction we find
being of the order of 2/3 in agreement with Maxted et al. (2001). Most
of the binaries found seemed to have formed via the first and second
CE ejection phases but we are starting to find systems that might have
formed through the first RLOF as predicted by Han et al. (2003).

\acknowledgements

The INT is operated on the island of La Palma by the ING in the
Spanish Observatorio del Roque de los Muchachos of the IAC. We thank
PATT for their support of this program. The Radcliffe telescope is
operated by the South African Astronomical Observatory.


\end{article}

\begin{thebibliography}{}

\bibitem[\protect\citeauthoryear{Cumming, Marcy \&\ Butler}{1999}]{cmb99}
Cumming A., Marcy G. W., Butler R. P., 1999, ApJ, 526, 890


\bibitem[\protect\citeauthoryear{Green, Liebert \&\
    Saffer}{2001}]{gls01} Green E. M., Liebert J., Saffer R. A., 2001,
    ASP Conf. Ser. Vol. 226, XII European Workshop on WDs

\bibitem[\protect\citeauthoryear{Han et al.}{2002}]{h02} Han Z. et
  al., 2002, MNRAS, 336, 449

\bibitem[\protect\citeauthoryear{Han et al.}{2003}]{h03} Han Z. et
  al., 2003, MNRAS, 341, 669

\bibitem[\protect\citeauthoryear{Heber, Reid \&\ Werner}{2000}]{hrw00}
  Heber U., Reid I. N., Werner K., 2000, A\&A, 363, 198

\bibitem[\protect\citeauthoryear{Marsh, Dhillon \&\
    Duck}{1995}]{mdd95} Marsh T. R., Dhillon V. S., Duck S. R., 1995,
    MNRAS, 275, 828

\bibitem[\protect\citeauthoryear{Maxted et al.}{2001}]{m01} Maxted
  P. F. L. et al., 2001, MNRAS, 326, 139


\bibitem[\protect\citeauthoryear{Maxted et al.}{2002}]{m02} Maxted
  P. F. L. et al., 2002, MNRAS, 333, 231

\bibitem[\protect\citeauthoryear{Morales-Rueda et al.}{2003}]{m03}
  Morales-Rueda L. et al., 2003, MNRAS, 338, 752

\bibitem[\protect\citeauthoryear{Napiwotzki}{1997}]{n97} Napiwotzki
  R., 1997, A\&A, 322, 256

\bibitem[\protect\citeauthoryear{Saffer et al.}{1994}]{s94} Saffer
  R. A. et al., 1994, ApJ, 432, 351

\bibitem[\protect\citeauthoryear{Tutukov \&\ Yungelson}{1990}]{tt90}
  Tutukov A., Yungelson L., 1990, SvA, 34, 57 

\end{thebibliography}
\end{document}